\documentclass[12pt,preprint]{aastex}

\shorttitle{3D MHD Simulations of Cold Fronts}
\shortauthors{Asai et al.}

\begin{document}

\title{Three-dimensional Magnetohydrodynamic Simulations of \\
Cold Fronts in Magnetically Turbulent ICM}

\author{N. Asai}
\affil{Graduate School of Science and Technology, 
Chiba University, 1-33 Yayoi-cho, Inage-ku, Chiba 263-8522, Japan}
\email{asai@astro.s.chiba-u.ac.jp}
\author{N. Fukuda}
\affil{Department of Computer Simulation,
Faculty of Informatics, Okayama University of Science,
1-1 Ridai-cho, Okayama 700-0005, Japan}
\email{fukudany@sp.ous.ac.jp}
\and
\author{R. Matsumoto}
\affil{Department of Physics, Faculty of Science, 
Chiba University, 1-33 Yayoi-cho, Inage-ku, Chiba 263-8522, Japan}
\email{matumoto@astro.s.chiba-u.ac.jp}

\begin{abstract}
Steep gradients of temperature and density, called cold fronts, are
 observed by {\it Chandra} in a leading edge of subclusters moving
 through the intracluster medium (ICM). The presence of cold fronts
 indicates that thermal conduction across the front is suppressed by
 magnetic fields. We carried out three-dimensional magnetohydrodynamic
 (MHD) simulations including anisotropic thermal conduction of a
 subcluster moving through a magnetically turbulent ICM. We found that
 turbulent magnetic fields are stretched and amplified by shear flows
 along the interface between the subcluster and the ambient ICM. Since
 magnetic fields reduce the efficiency of thermal conduction across the
 front, the cold front survives at least $1 \,{\rm Gyr}$. We also found
 that a moving subcluster works as an amplifier of magnetic
 fields. Numerical results indicate that stretched turbulent magnetic
 fields accumulate behind the subcluster and are further amplified by
 vortex motions. The moving subcluster creates a long tail of ordered
 magnetic fields, in which the magnetic field strength attains
 $\beta=P_{\rm gas}/P_{\rm mag}\la 10$. 
\end{abstract}


\section{Introduction}\label{int}
X-ray observations of clusters of galaxies by {\it ASCA} revealed
complex temperature distributions in the intracluster medium (ICM), in
which hot and cool plasma coexists (e.g, Perseus Cluster,
\citealt{arn94,fur01}). Sharp discontinuities of density and temperature
in ICM, called cold fronts, were found by high spatial resolution
observations by {\it Chandra} \citep{mar00, vik01b}. Cold fronts
manifest the coexistence of hot and cool plasma in clusters of galaxies.
They provide a key to understand thermal properties of the ICM.

Cold fronts in merging clusters such as A2142, A3667, and 1E0657-56
result from merging. When a subcluster is moving in the ICM, a sharp
boundary is formed between the subcluster and the ambient hot ICM in the
forehead of the subcluster because the cold plasma confined by the
subcluster is subjected to the ram pressure. The cold fronts are
not shock fronts because the {\it Chandra} images of the X-ray surface
brightness show that the temperature decreases on the denser side.  
A3667 has a clear, large-scale ($\sim 500\,{\rm kpc}$), arc-shaped cold
front which shows a steep gradient of the X-ray surface brightness and
temperature. The temperature decreases toward the denser part from
$8 \,{\rm keV}$ to $4\,{\rm keV}$ within $5 \, {\rm kpc}$. This
thickness of the front is 2-3 times smaller than the Coulomb mean free
path \citep{vik01b}.

A question which needs to be answered is how the temperature gradient is
created and sustained in  the ICM which typically has high Spitzer
conductivity, \\
$\kappa_{\rm Sp}= 5 \times 10^{-7} T^{5/2} \, {\rm erg \, s^{-1} \,
cm^{-1} \, K^{-1}}$ \citep{spi62}. 
Thermal conduction rapidly smooths such a steep gradient (e.g.,
\citealt{tak77}). The time required for heat to diffuse by conduction
across a length $L$ in the ICM is roughly given by 
$\tau_{\rm Sp} \sim \rho L^{2}/ \kappa_{\rm Sp} 
\sim 10^{7} (kT/ 5 \, {\rm keV})^{-5/2} (n / 10^{-3} \, 
{\rm cm^{-3}})(L/100 \, {\rm kpc})^{2} \, {\rm yr}$, 
where $\rho$ is the density. 
\citet{ett00} and \citet{mar03} estimated that the effective thermal
conduction is at least an order of magnitude lower than the Spitzer
value. 
This suggests that the thermal conduction across the front is suppressed
by magnetic fields parallel to the front \citep{vik01b}. \citet{vik01a}
suggested that ordered magnetic fields are formed in front of the
subcluster because small-scale turbulent magnetic fields are compressed
and stretched along the front ahead of the subcluster by its motion. 
Typical clusters of galaxies possess magnetic fields of 
$\sim \,\mu {\rm G}$ (e.g., \citealt{kro94,car02}).
\citet{joh04} reported that $1-2\,\mu {\rm G}$ fields, tangled on 
$100\,{\rm kpc}$ scale pervade the central region of A3667. When
magnetic fields exist, the characteristic scale of the heat exchange
across the field lines is reduced significantly compared with
non-magnetized ICM to the Larmor radius, 
$r_{\rm L}\sim \,2500\, (B/1\,\mu {\rm G})^{-1}\,(T/5 \,{\rm keV})^{1/2} \, {\rm km}$.
Intracluster magnetic fields play a crucial role for thermal conduction
even if the magnetic pressure is lower than the gas pressure. 

A number of authors reported that cold fronts were reproduced in
numerical simulations as a result of a merging process 
(e.g., \citealt{bia02,nag03,hei03,acr03,tak05}). 
In these simulations, however, magnetic fields and thermal conduction
were not included. To study the evolution of intracluster magnetic
fields, \citet{roe99} performed MHD simulations of merging clusters and
\citet{dol02} performed cosmological MHD simulations. However, they
ignored the thermal conduction.
\citet{asa04} performed two-dimensional (2D) MHD simulations of a
subcluster moving through uniform magnetic fields by including
anisotropic thermal conduction. They showed that ordered magnetic fields
wrap the subcluster and suppress the thermal conduction across the front. 
This work was extended to three-dimension by \citet{asa05}.

Some authors pointed out that the effective conductivity in turbulent
magnetic fields is only several times lower than the Spitzer value
\citep{nar01}. \citet{dol04} carried out cosmological
hydrodynamic simulations including the thermal conduction with the
isotropic effective conductivity, $\kappa \sim \kappa_{\rm Sp}/3$ and
showed that temperature gradients are smoothed compared to the case
without thermal conduction. However, a moving subcluster may stretch the
turbulent magnetic fields and such fields may reduce the conductivity.

Lyutikov (2006) studied the magnetic draping mechanism by which a
strongly magnetized, thin boundary layer with a tangential magnetic
field is created in the boundary between a moving cloud and the ambient
plasma. He theoretically predicted that for supersonic cloud motion,
magnetic field strength inside the layer reaches near equipartition
values with thermal pressure.

\citet{asa04} carried out a simulation starting from a disordered
magnetic field and showed that cold fronts can be sustained because
magnetic fields are stretched along the front and wrap the
subcluster. However, their simulation was limited to 2D and their
initial magnetic field still had a large coherent length.
Thus magnetic fields can easily suppress the thermal conduction across
the front.

In this paper, we present results of three-dimensional (3D) MHD
simulations of a subcluster moving through turbulent magnetic fields to
show that even when magnetic fields are turbulent, cold fronts can be
sustained. The paper is organized as follows. In \S \ref{model}, we
describe the initial condition and model parameters. Simulation results
are presented in \S \ref{res}: we show the effects of magnetic fields on
cold fronts and amplification of magnetic fields behind the subcluster. 
Finally, we discuss and summarize our results in \S \ref{sum}. 

\section{Simulation Model}\label{model}
We simulated the time evolution of a cluster plasma in a frame comoving
with the subcluster. The basic equations are as follows:
\begin{equation}
\frac{\partial \rho}{\partial t} + \mbox{\boldmath $\nabla$ $\cdot$}
 (\rho \mbox{\boldmath $v$}) = 0,
\label{eq1}
\end{equation}
\begin{equation}
\rho \left[
\frac{\partial \mbox{\boldmath $v$}}{\partial t} 
+ (\mbox{\boldmath $v$ $\cdot$ $\nabla$) $v$}
\right] =
-\mbox{\boldmath $\nabla$}{\it p} + 
\frac{(\mbox{\boldmath $\nabla$ $\times$ $B$) $\times$ $B$}}{4 \pi} 
- \rho \mbox{\boldmath $\nabla$} \psi, 
\label{eq2}
\end{equation}
\begin{equation}
\frac{\partial \mbox{\boldmath $B$}}{\partial t} 
= \mbox{\boldmath $\nabla$ $\times$ ( $v$ $\times$ $B$)},
\label{eq3}
\end{equation}
\begin{equation}
\frac{\partial}{\partial t} 
\left[
\frac{1}{2} \rho v^{2} 
+ \frac{B^{2}}{8 \pi} + \frac{p}{\gamma -1}
\right]
+ \mbox{\boldmath$\nabla$$\cdot$}
\left[
\left(\frac{1}{2} \rho v^{2} + \frac{\gamma p}{\gamma -1}
\right) \mbox{\boldmath$v$} + 
\frac{(- \mbox{\boldmath $v$ $\times$ $B$) $\times$ $B$}}{4 \pi}  - \kappa_{\parallel}  
\mbox{\boldmath $\nabla_{\parallel}$} T
\right] 
= - \rho \mbox{\boldmath $v$ $\cdot$ $\nabla$} \psi,
\label{eq4}
\end{equation}
where $\rho$, $\mbox{\boldmath $v$}$, $p$, $\mbox{\boldmath $B$}$, and
$\psi$ are the density, velocity, pressure, magnetic fields, and
gravitational potential, respectively. We use the specific heat ratio 
$\gamma = 5/3$. The subscript $\parallel$ denotes the components
parallel to the magnetic field lines. We assume that heat is conducted
only along the magnetic field lines. 
We solved equations (\ref{eq1})-(\ref{eq4}) in a Cartesian coordinate
system $(x, \,y, \,z)$ by using a solver based on a modified
Lax-Wendroff method \citep{rub67} with artificial viscosity 
\citep{ric67} implemented to the Coordinated Astronomical Numerical 
Software (CANS). We did not include the physical viscosity.
Artificial viscosity is included only in regions close to the
discontinuities to suppress numerical oscillations. It does not affect
the dynamics in smooth regions.
The thermal conduction term in the energy equation is solved by the
implicit red and black successive over-relaxation method 
\citep[see][for detail]{yok01}. 
The radiative cooling term is not included. The units of length,
velocity, density, pressure, temperature, and time in our simulations
are
$r_{0}=250 \, {\rm kpc}$,
$v_{0}=1000 \, {\rm km \, s^{-1}}$,
$\rho_{0}=5 \times 10^{-27} \, {\rm g \, cm^{-3}}$,
$p_{0} = 3 \times 10^{-11} \, {\rm erg \, cm^{-3}}$,
$kT_{0}= 4 \, {\rm keV}$, and
$t_{{0}}=r_{0}/v_{0}= 3 \times 10^{8} \, {\rm yr}$,
respectively.

We carried out simulations for 6 models. Table \ref{tbl-1} shows the
model parameters. When magnetic fields exist, heat conducts only
parallel to the magnetic field lines. Model MT1 and MT2 are models with
turbulent magnetic fields. We define Fourier components of magnetic
vector potential,
$\mbox{\boldmath $\tilde{A}$} (k) = \mbox{\boldmath $\tilde{A}$}_{0} k^{-\alpha}$,
where $k$ is a wave number. The amplitudes 
$\mbox{\boldmath $\tilde{A}$}_{0}$ are taken to be random by using
random numbers, and we adopt $\alpha =5/3$. This vector potential 
$\mbox{\boldmath $\tilde{A}$}$ in $k$-space is transformed to vector
potential $\mbox{\boldmath $A$}$ in physical space via a 3D fast Fourier
transform (FFT). We computed a tangled divergence-free initial magnetic
field via $\mbox{\boldmath $B$} =\mbox{\boldmath $\nabla$ $\times$ $A$}$.
On the other hand, models MU1 and MU2 are models with uniform magnetic 
fields parallel to the $z$-direction and perpendicular to the motion 
of the subcluster.

\begin{figure}[ht]
\epsscale{.60}
\plotone{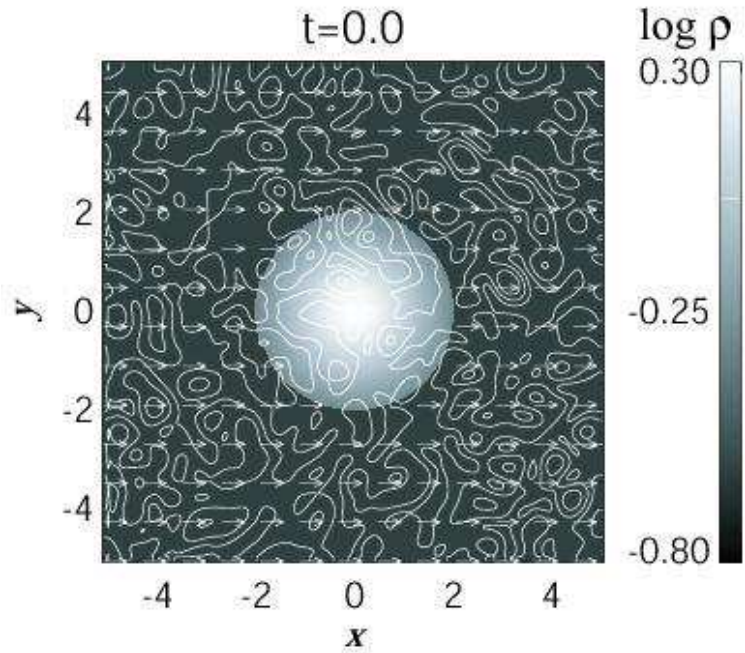}
\caption{Initial distribution of logarithm of density at $z=0$
 plane. Solid curves and arrows show the contours of strength of
 magnetic fields and velocity vectors, respectively.
\label{fig1}}
\end{figure}

Figure \ref{fig1} shows the initial density distribution for model MT1
at $z=0$ plane. Solid curves and arrows show the contours of magnetic
field strength and velocity vectors, respectively. We assume that a
spherical isothermal low-temperature ($kT_{\rm in} = 4 \, {\rm keV}$)
plasma is confined by the gravitational potential of the subcluster. The
subcluster has a $\beta$-model density distribution,
\begin{equation}
\rho_{\rm in}=\rho_{\rm c}
\left[
1+
\left(
\frac{r}{r_{\rm c}}
\right)^{2}
\right]^{-3 \beta' /2}
\end{equation}
where we adopted $\beta' = 2/3$, the core radius 
$r_{\rm c}=290\, {\rm kpc}$, and the maximum density 
$\rho_{c}=2\,\rho_{0}= 10^{-26} \, {\rm g \, cm^{-3}}$.
The subcluster is embedded in the low density 
($\rho_{\rm out}=\rho_{0}/4$), hot ($kT_{\rm out}=2\,kT_{\rm in}$)
ambient plasma. We assume that the subcluster is initially in
hydrostatic equilibrium and has a jump
of density and temperature with respect to the ambient plasma.
We also assume that the ambient plasma has a uniform speed 
$M= v_{x}/c_{\rm s,\,out}=1$, where $c_{\rm s, \,out}$ is the ambient
sound speed.
Note that magnetic fields exist even inside the subcluster in all models with 
magnetic fields.
The box size of our simulations is 
$(2.5\, {\rm Mpc})^{3}$. We used $256^{3}$ grid points for typical
models. The numerical resolution is $\sim 10\,{\rm kpc}$.

An important parameter is plasma $\beta$ defined as the ratio of the gas
pressure to the magnetic pressure (see Table \ref{tbl-1}). 
The initial mean field strength $B_{0}$ is $\sim 0.03\, \mu {\rm G}$ for 
model MT1 and $\sim 0.09 \, \mu {\rm G}$ for model MT2.
Correspondingly, initial plasma $\beta$ for these models is 
$\beta_{0} \sim 7.5\times 10^{4}$ and $\sim 8.3 \times 10^{3}$, respectively.
For models with uniform fields, the strength $B_{0}$ is $\sim 0.07\, \mu {\rm G}$ 
for models MU1 and $\sim 0.27 \, \mu {\rm G}$
for model MU2. The initial plasma $\beta$ for these models is 
$\beta_{0} \sim 1.0 \times 10^{4}$ and $\sim 1.0 \times 10^{3}$,
respectively. Model MT3 is the same as model MT1 except that thermal
conduction is ignored and the box size of the $x$-direction is $1.5$
times larger than that of model MT1. We also carried out a simulation
for a model without magnetic fields (model H), including isotropic
thermal conduction.

For turbulent field models (MT1, MT2, and MT3), the left boundary at
$x=-5$ is taken to be the fixed boundary, except for the magnetic
fields. The magnetic fields at the left boundary are extracted from the
initial distribution of magnetic fields as follows, 
\begin{eqnarray}
\mbox{\boldmath $B$}(1,\,j,\,k)=\mbox{\boldmath $B$}_{0}(ix -v_{x {\rm 0}}\, t,\,j,\,k),
\end{eqnarray}
where $B_{0}$ is the initial magnetic field, $j$ and $k$ are mesh
numbers in the $y$ and $z$ directions, $ix$ is the total number of grid
points in the $x$-direction, $v_{x {\rm 0}}$ is the initial velocity of
the subcluster, and $t$ is the time. For other models, the left boundary
is taken to be a fixed boundary. In all models, boundaries other than
the left boundary are free boundaries where waves can be transmitted.

\section{Results}\label{res}
\subsection{Time Evolution of a Subcluster and Magnetic Fields}\label{res1}
\begin{figure}[ht]
\epsscale{.90}
\plotone{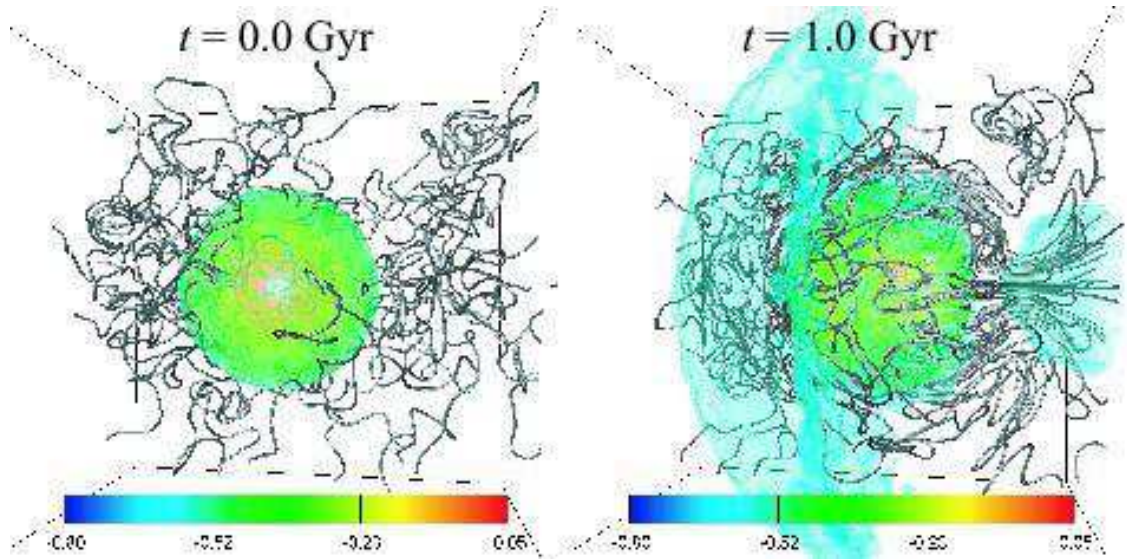}
\caption{Time evolution of density distribution and magnetic field
 lines. Left and right panels show the distributions of density and
 magnetic field lines  at $t=0.0\, {\rm Gyr}$ and $t=1.0 \, {\rm Gyr}$,
 respectively. Color shows isosurfaces of density (${\rm log}\, \rho$). 
 Solid curves show magnetic field lines.
\label{fig2}}
\end{figure}

Figure \ref{fig2} shows time evolution of density distribution and
magnetic fields for model MT1. The left panel shows the initial state
and the right panel shows the distribution at $t=1.0 \,{\rm Gyr}$.
Solid curves show the magnetic field lines. Since the subcluster plasma
moves with sound speed, a bow shock appears ahead of the subcluster.
Magnetic field lines are stretched along the subcluster surface at
$t=1.0\,{\rm Gyr}$ due to the ambient gas motion. Thus, the motion of
the subcluster creates ordered magnetic fields along the interface
between the subcluster and ambient ICM. The magnetic fields accumulate
behind the subcluster.

\subsection{Effect of Magnetic Fields on Thermal Conduction}\label{res2}
\begin{figure}[ht]
\epsscale{.95}
\plotone{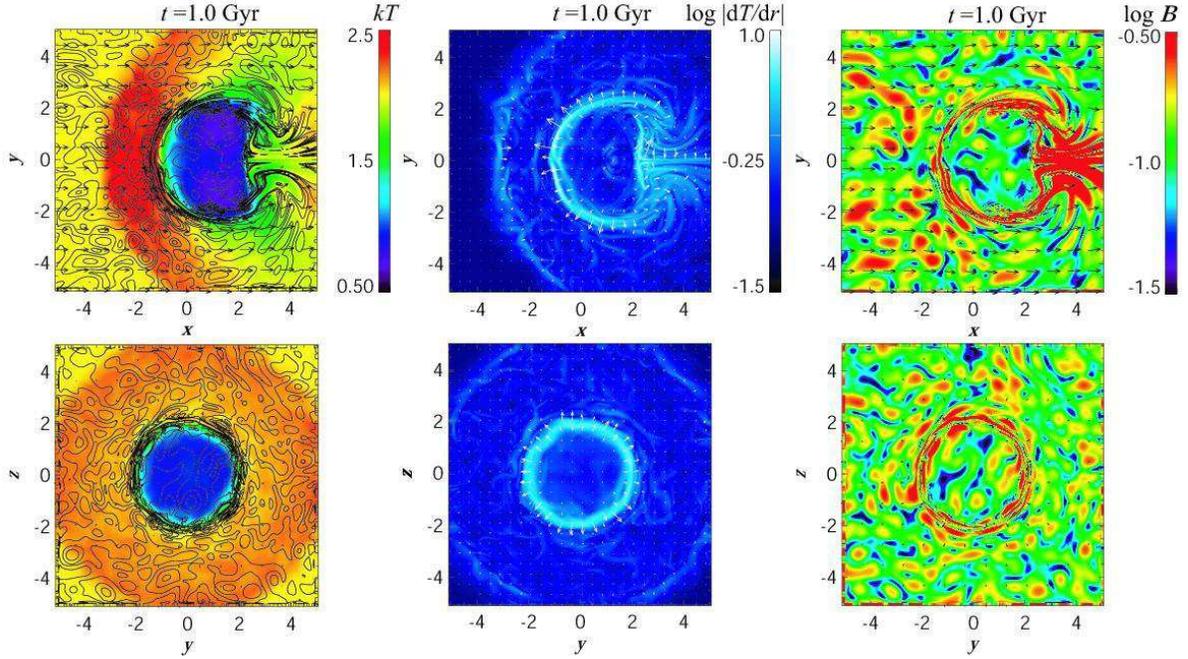}
\caption{Distributions of temperature (left panels), gradients of
 temperature (middle panels), and magnetic field strength (right panels)
 at $t = 1.0 \, {\rm Gyr}$ for model MT1. Upper panels show the slice at
 $z = 0$ and lower panels show the slice at $x = 0$. Solid curves in
 left panels show the contours of magnetic field strength. Arrows in the
 left and right panels show the velocity vectors. Arrows in the middle
 panels show the gradients of temperature.
\label{fig3}}
\end{figure}

Figure \ref{fig3} shows snap shots of distributions of temperature (left
panels), temperature gradients (middle panels), and magnetic field
strength (right panels) at $t = 1.0 \, {\rm Gyr}$. The upper panels show
the slices at $z = 0$ plane and the bottom panels show the slices at $x
= 0$ plane for model MT1, respectively. Solid curves in the left panels
are the contours of magnetic field strength. Arrows in the left (and
right) panels show the velocity vectors, and those in the middle panels
are the gradients of temperature.

The temperature distributions (left panels) show that steep temperature 
gradients are maintained at $t=1.0 \,{\rm Gyr}$ because the thermal
conduction across the front is suppressed by stretched magnetic field
lines wrapping the subcluster. The middle panels show that steep
temperature gradients around the subcluster surface are sustained. The
cold front is located at $(x,\,y)\sim (-1,\,0)$ in $z=0$ plane (upper
panels). We can also identify the bow shock at $(x,\,y)\sim(-3,\,0)$ in
$z=0$ plane. The steepest temperature gradient is seen in the vicinity
of $(x,\, y)=(-1,\, 0)$ in the upper middle panel. The distributions in
$x=0$ plane (bottom panels) clearly show that the subcluster is almost
entirely covered with magnetic fields. 

In the left and right panels of Figure \ref{fig3}, we can see that
magnetic fields are stretched and compressed around the subcluster as we
already mentioned. A shear flow along the boundary between the
subcluster and the ambient plasma stretches magnetic fields. Moreover,
magnetic field strength is amplified behind the subcluster 
(see \S \ref{res3} for details). The field amplification is more
prominent in the tail of the subcluster than in the forehead. The
amplification of magnetic fields ahead of the cold fronts is due to the
compression of magnetic fields by the ambient plasma flow
hitting the subcluster. The ambient plasma flowing along the subcluster
surface converges to the $x$-axis behind the subcluster. Therefore,
magnetic fields frozen to the plasma accumulates behind the subcluster. 

\begin{figure}[ht]
\epsscale{.90}
\plotone{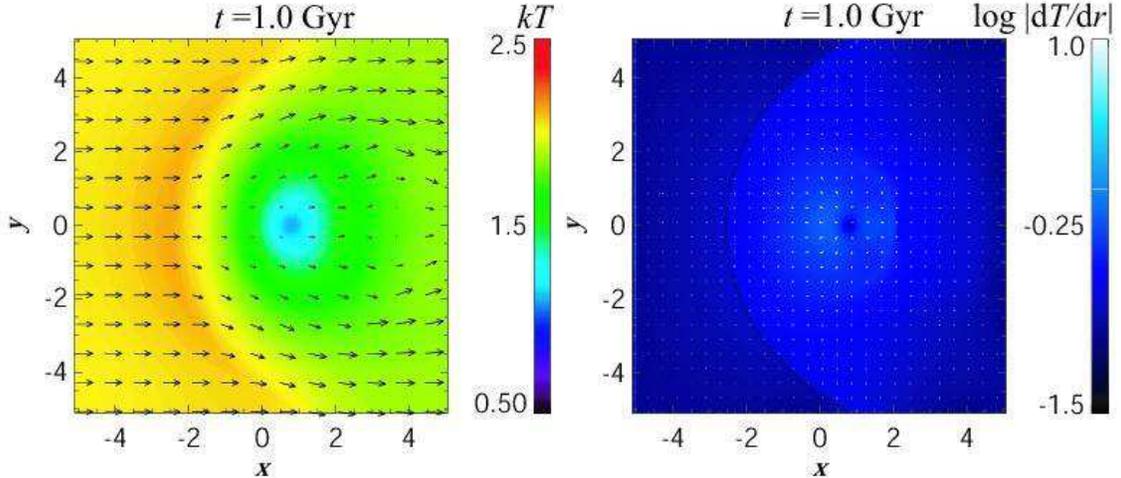}
\caption{Distributions of temperature (left) and gradients of
 temperature  (right) in $z=0$ plane at $t = 1.0 \, {\rm Gyr}$ in
 hydrodynamic model (model H). Arrows show the velocity vectors (left)
 and the gradients of temperature (right), respectively.
\label{fig4}}
\end{figure}

In order to demonstrate the effects of magnetic fields on thermal
conduction, we present results of hydrodynamic model (model H) in Figure
\ref{fig4}. The left and right panels show the distributions of
temperature and temperature gradients. Compared with the results shown in
Figure \ref{fig3}, temperature gradients are smeared out because
isotropic thermal conduction from the ambient hot plasma rapidly heats
up the dense cool plasma confined in the subcluster. 

\begin{figure}[ht]
\epsscale{.95}
\plottwo{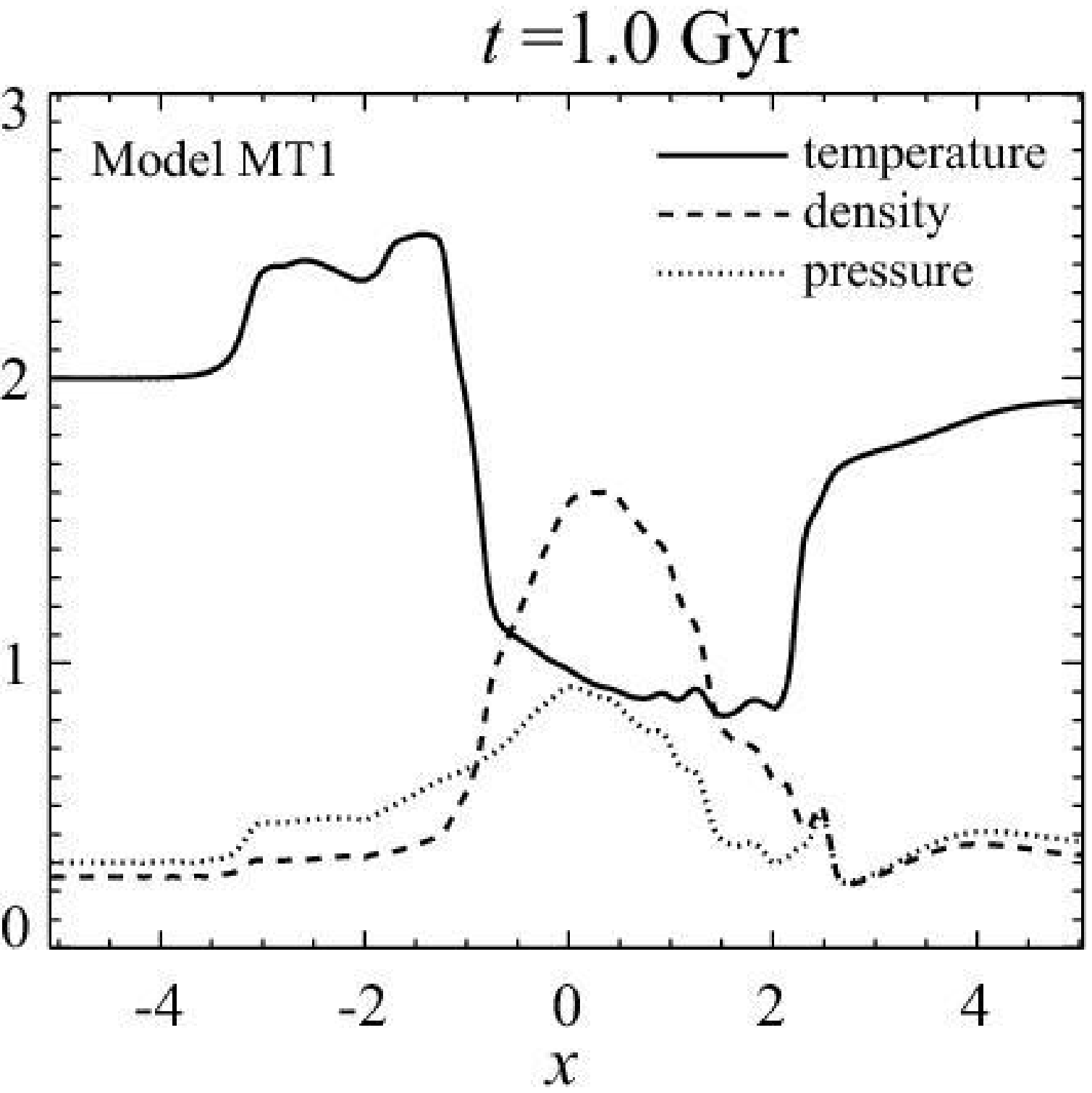}{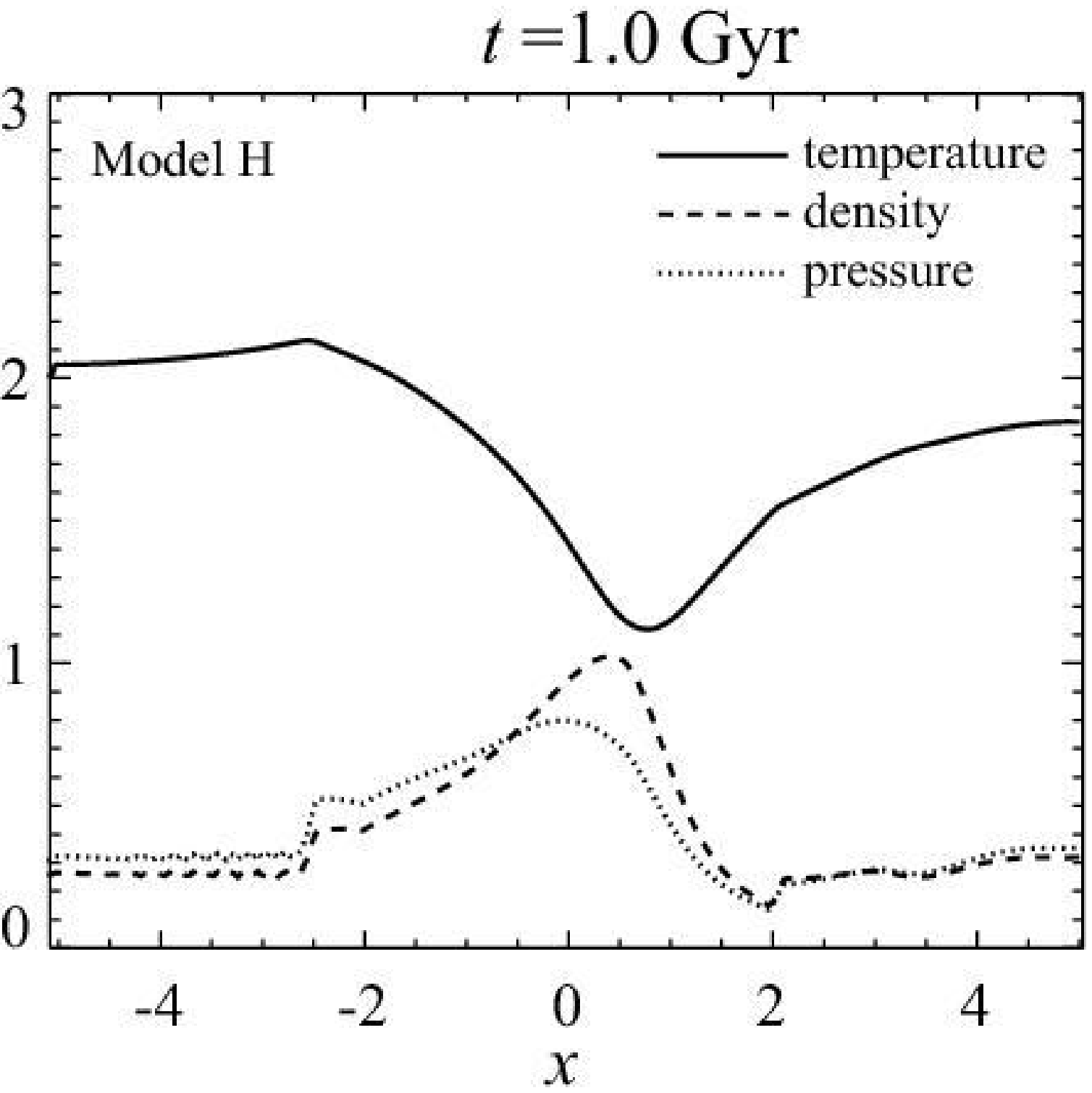}
\caption{Comparison of distributions of temperature, density, and
 pressure along $x$-axis ($y=z=0$) at $t=1.0\, {\rm Gyr}$ for models MT1
 (left) and H (right). Solid, dashed, and dotted curves show the
 temperature, density, and pressure, respectively. 
\label{fig5}}
\end{figure}

In Figure \ref{fig5}, we show the distributions of quantities at 
$t=1.0 \, {\rm Gyr}$ along the $x$-axis ($y=z=0$). The left and right
panels show the distributions of temperature (solid curve), density
(dashed curve), and pressure (dotted curve) for models MT1 and H. The
temperature distribution in the left panel shows that a steep gradient
exists at $x=-1$, while pressure distribution is smooth. This feature is
consistent with the observed features of cold fronts (e.g.,
\citealt{mar00}). On the other hand, when magnetic fields do not exist
(right panel), the subcluster plasma is subjected to the isotropic
thermal conduction. After $t=1.0 \, {\rm Gyr}$, the subcluster
evaporates because of the conduction from the ambient hot plasma. The
peak density in the subcluster becomes lower than that in the initial
state. 

\subsection{Amplification of Magnetic Fields}\label{res3}
\begin{figure}[ht]
\epsscale{.95}
\plottwo{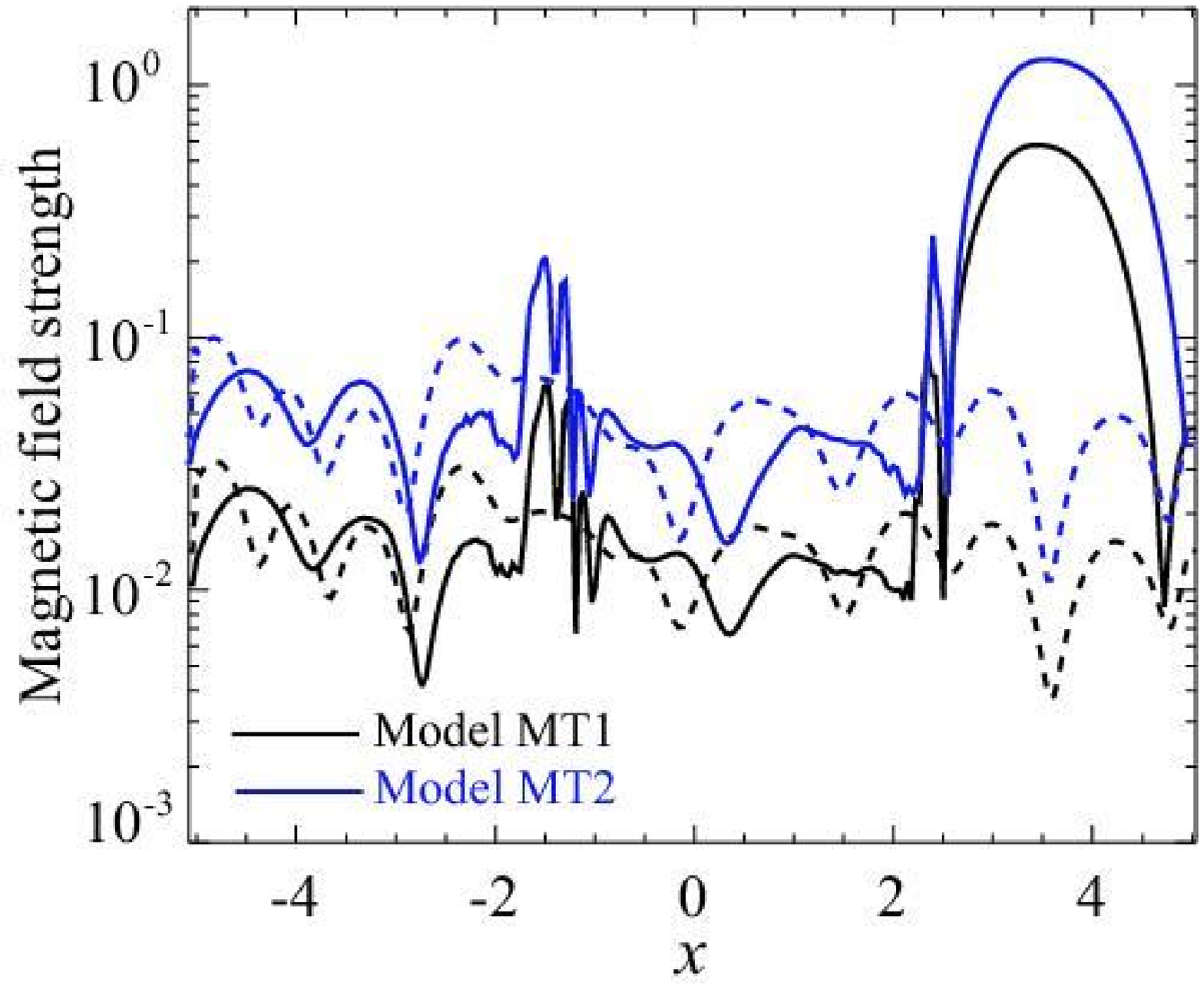}{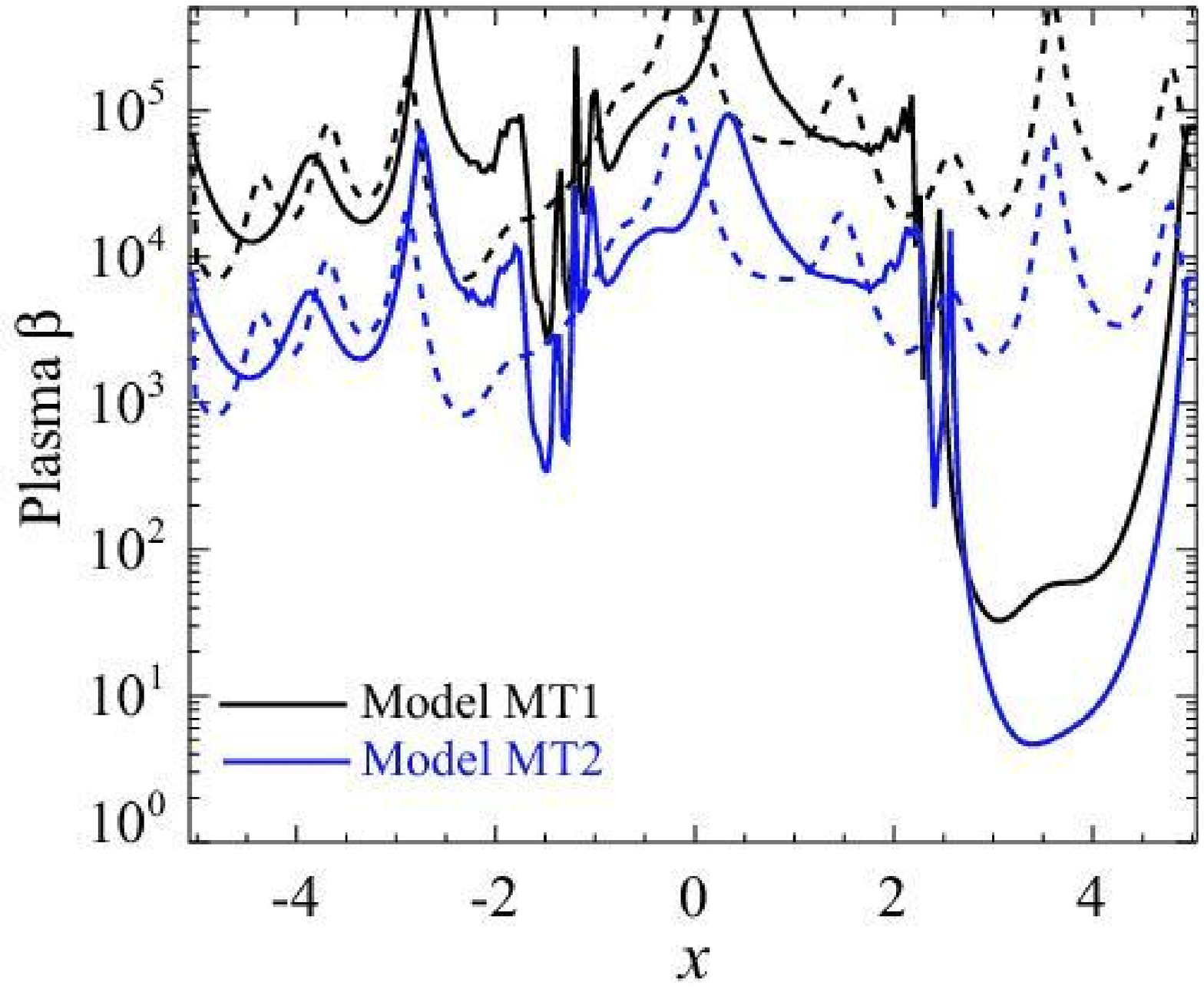}
\caption{Distributions of magnetic field strength (left) and plasma
 $\beta$ (right) for models MT1 and MT2. Black curves in both panels
 show the distributions along the $x$-direction, 
$(y,\,z)=(-0.04, \,-0.20)$ for model MT1. Blue curves in both panels
 show the distributions along the $x$-direction, 
$(y,\,z)=(-0.04, \,-0.24)$ for model MT2. Solid and dashed curves in
 both panels show those at $t=1.0\,{\rm Gyr}$ and the initial state,
 respectively.
\label{fig6}}
\end{figure}

Simulation results revealed other interesting features. We found that a
moving subcluster works as an amplifier of magnetic fields. The left
panel of Figure \ref{fig6} shows the distributions of magnetic field
strength for the turbulent field models (models MT1 and MT2). The right
panel shows the distributions of plasma $\beta$. We plot these
distributions along $(y,\,z)=(-0.04, \,-0.20)$ for model MT1 (black
curves) and $(y,\,z)=(-0.04, \,-0.24)$ for model MT2 (blue curves),
respectively. Solid curves in both panels are distributions at 
$t=1.0 \,{\rm Gyr}$ and dashed curves show those at the initial state,
respectively. When turbulent fields exist, field strength is amplified
in front of the subcluster and behind it. The amplification of the field
strength is most prominent behind the subcluster. The field strength in
both models increases about 30 times with respect to the averaged
initial value. The right panel shows that plasma $\beta$ in both models
decreases. In model MT2, plasma $\beta$ decreases below $\beta \sim 10$
in the tail of the subcluster.

\begin{figure}[ht]
\epsscale{.95}
\plottwo{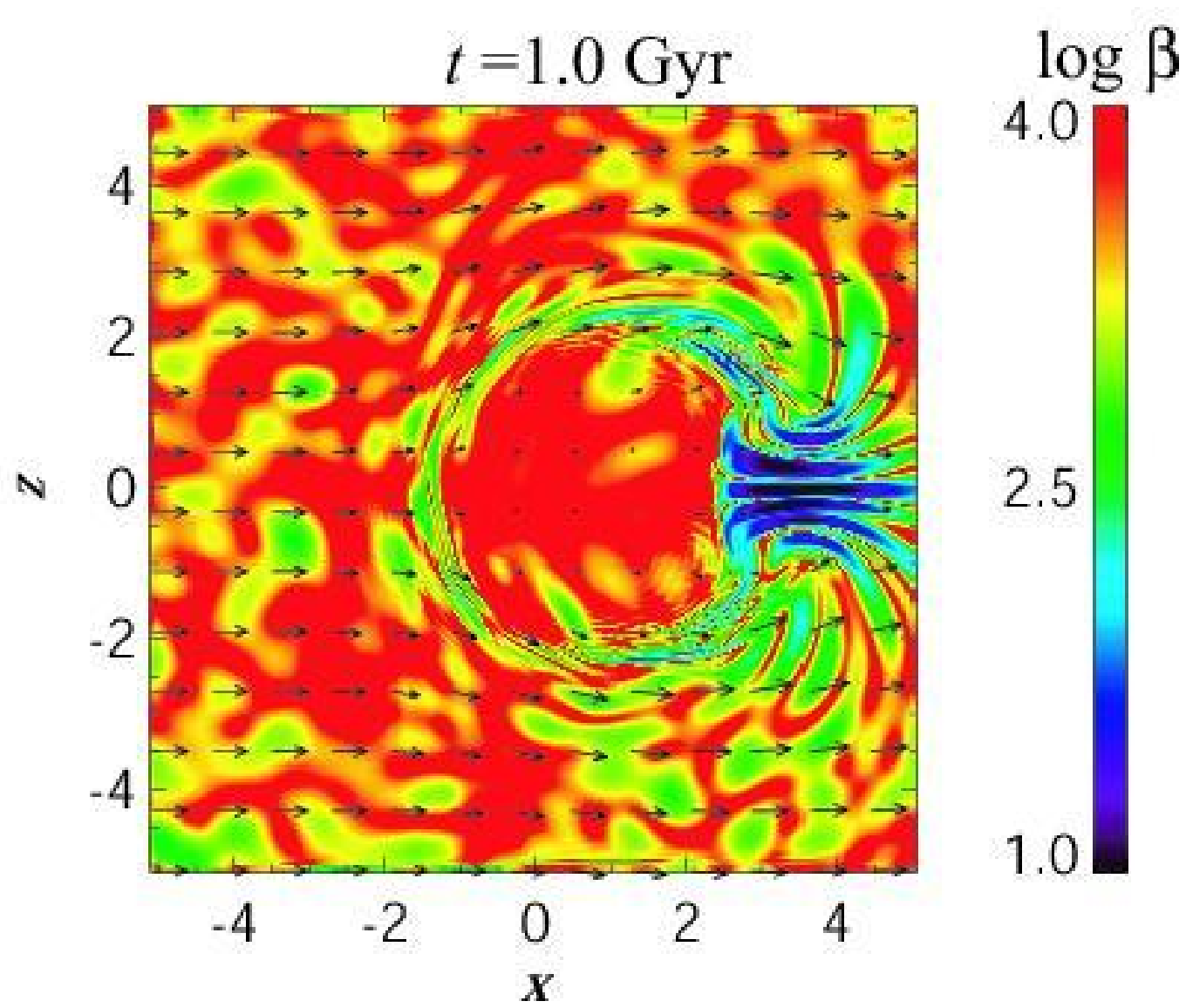}{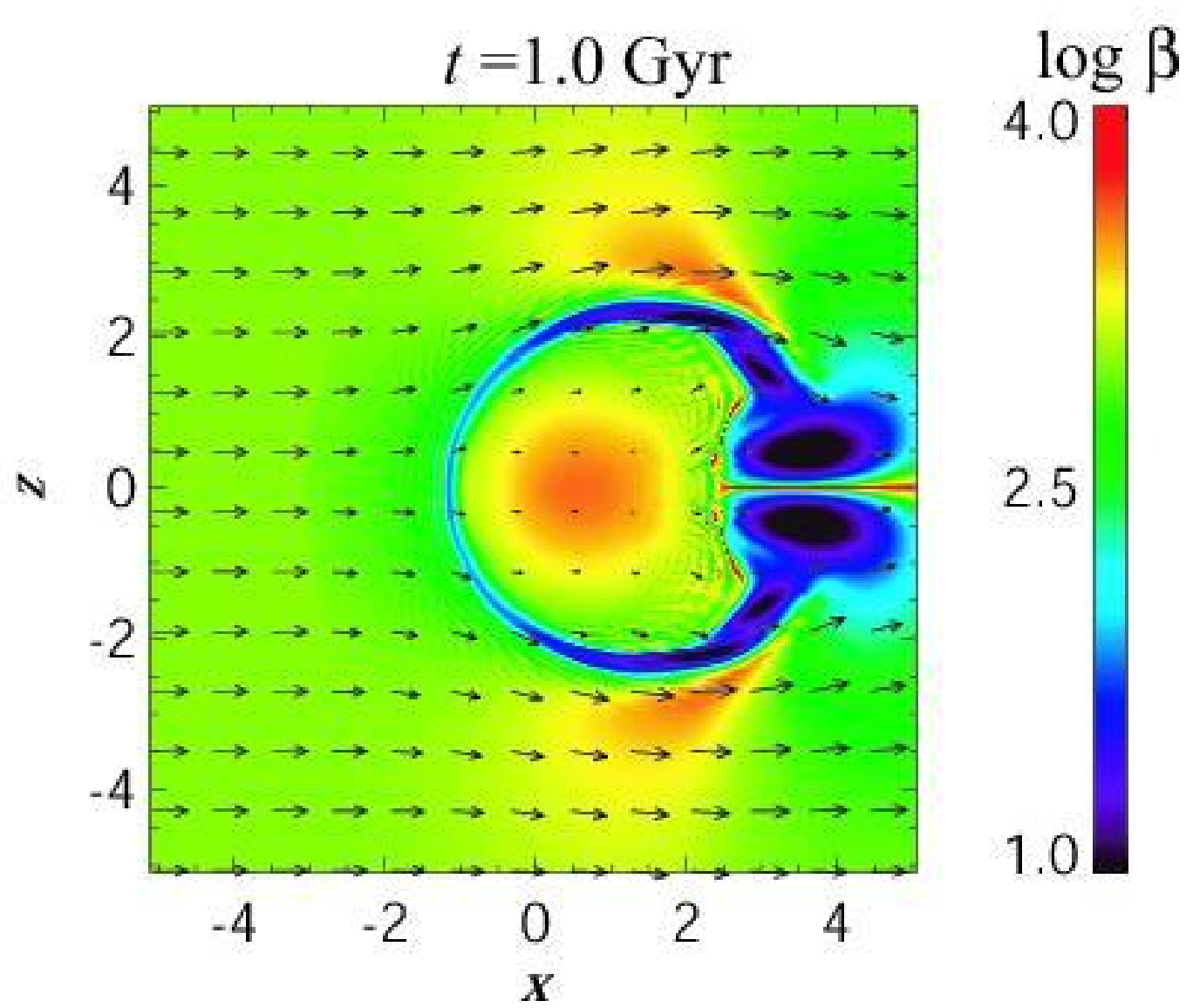}
\caption{Distributions of plasma $\beta$ in $y=0$ plane at $t = 1.0 \, {\rm Gyr}$ 
for model MT2 (left) and model MU2 (right). Arrows show the velocity
 vectors. \label{fig7}}.
\end{figure}

Let us compare the results for the turbulent field models with those of
uniform field models. Figure \ref{fig7} shows the distribution of plasma
$\beta$ at $t=1.0 \,{\rm Gyr}$ along $y=0$ plane for model MT2 (left)
and that for model MU2 (right). Note that the initial direction of
magnetic fields for model MU2 is parallel to the $z$-axis. In both
panels, plasma $\beta$ decreases remarkably behind the subcluster. The
plasma $\beta$ in this region is lower than $\beta \sim 10$. The region
of lower plasma $\beta$ is larger for model MU2 than that for model MT2
because the initial magnetic fields are uniform in model MU2, thus the
ambient plasma flow creates the ordered fields easily behind the
subcluster. An important finding is that plasma $\beta$ decreases to
$\beta \sim 10$ behind the subcluster even if the initial magnetic fields
are turbulent. 

\begin{figure}[ht]
\epsscale{.95}
\plottwo{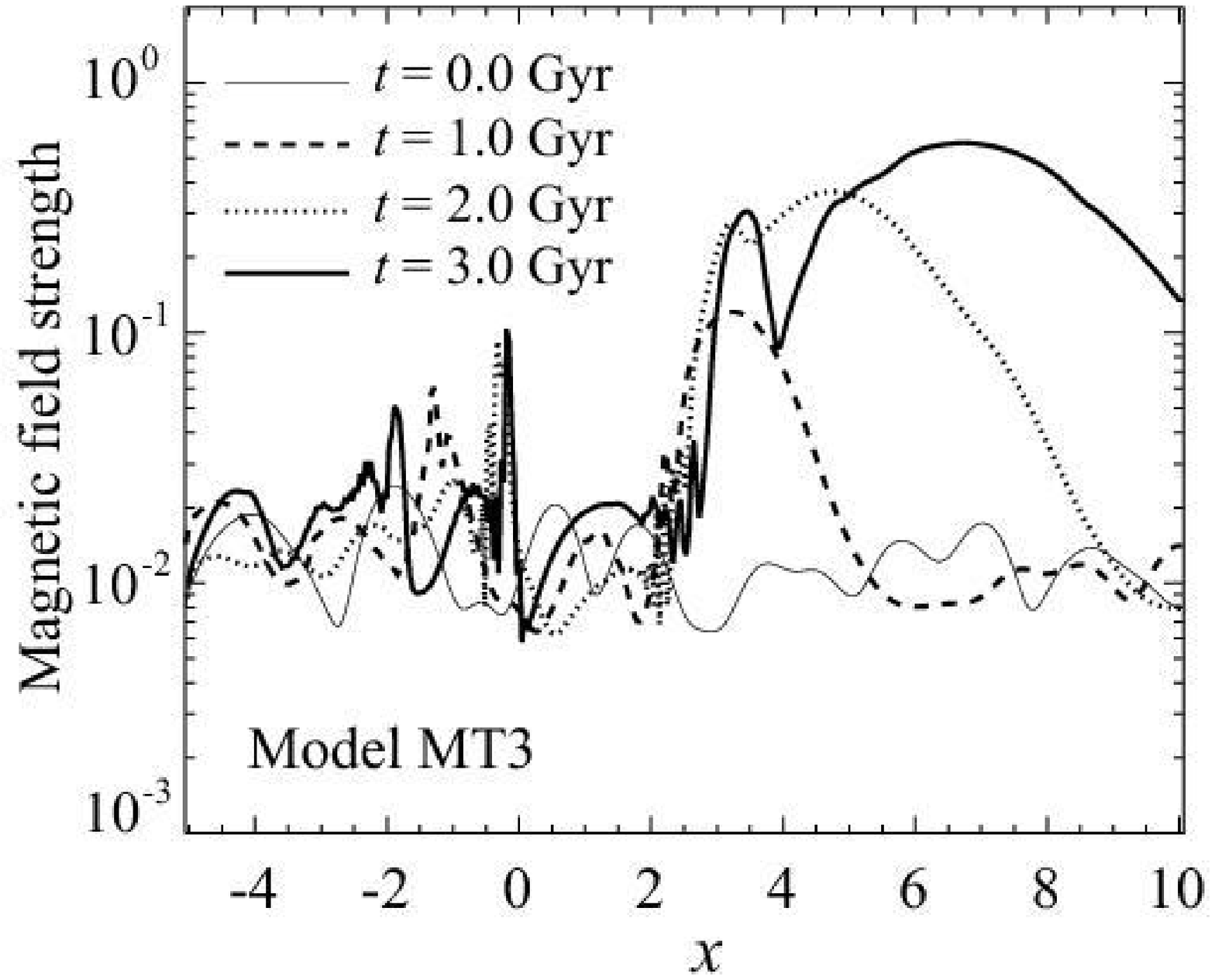}{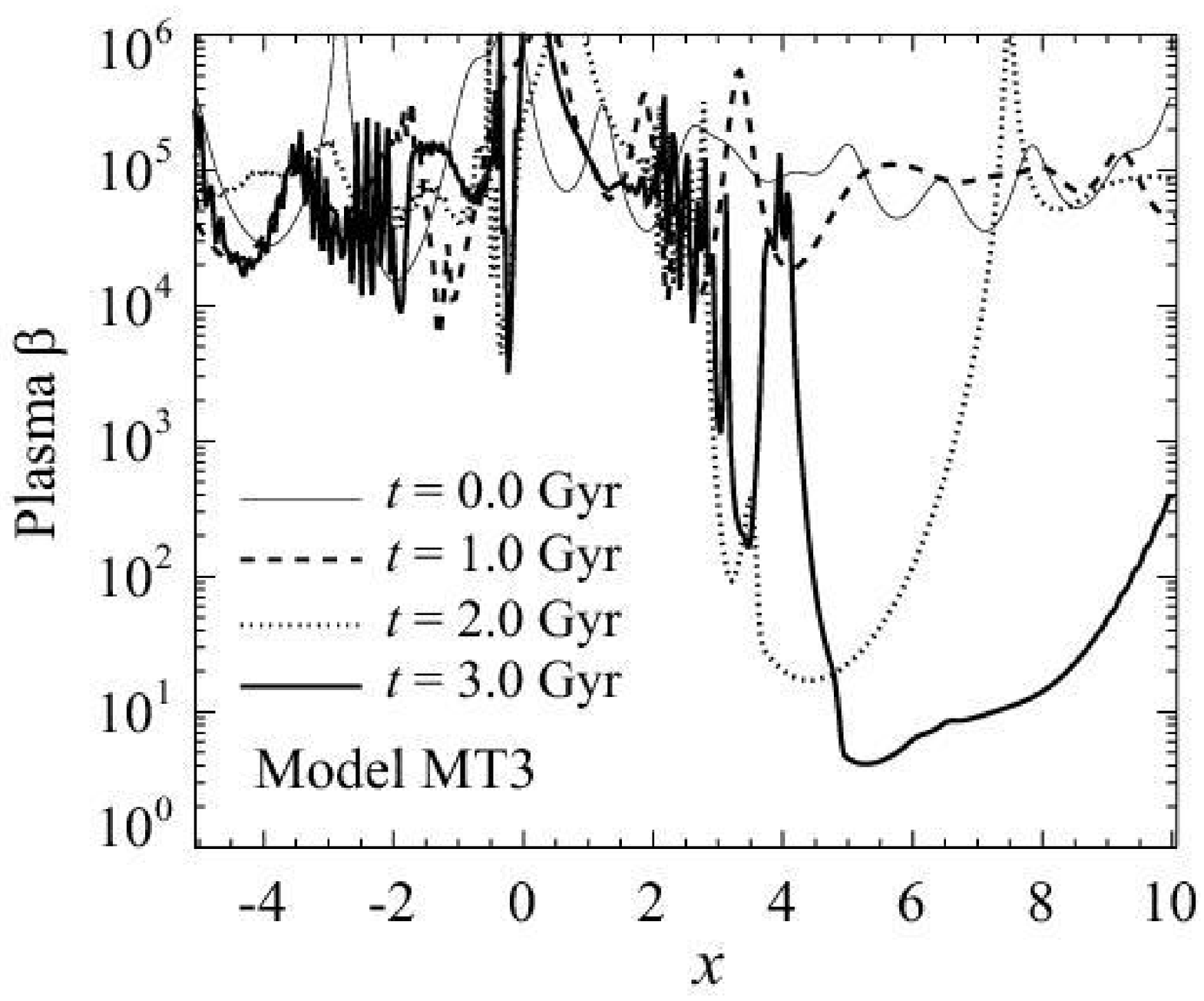}
\caption{Distributions of magnetic field strength (left) and plasma
 $\beta$ (right) for model MT3 along $x$-direction. We plot them along
 $y=z=-0.08$ for both panels. The length of the $x$-axis is $3.75 \,{\rm
 Mpc}$. Curves show the distributions for $t=0.0 \,{\rm Gyr}$ (thin
 solid curve), $1.0 \,{\rm Gyr}$ (dashed curve), $2.0 \,{\rm Gyr}$
 (dotted curve), and $3.0 \,{\rm Gyr}$ (thick solid curve),
 respectively.
\label{fig8}}
\end{figure}

In model MT3, we used larger simulation box and carried out a simulation
until $t=3.0 \,{\rm Gyr}$ in order to follow the growth of magnetic
fields behind the subcluster. 
The size of simulation box for model MT3 is $1.5$ times larger in the
$x$-direction than that for other models. Figure \ref{fig8} shows the
distributions of magnetic field strength (left) and plasma $\beta$
(right) along the $x$-direction ($y=z=-0.08$) for model MT3. Curves in
both panels show the distribution at $t=0.0 \, {\rm Gyr}$ (thin solid
curve), $1.0\,{\rm Gyr}$ (dashed curve), $2.0 \,{\rm Gyr}$ (dotted
curve), and $3.0\,{\rm Gyr}$ (thick solid curve). Magnetic fields are
amplified with time behind the subcluster. They are accumulated
around $x\sim 3$ at $t=1.0\,{\rm Gyr}$. The flow motion stretches
magnetic fields along the $x$-direction.
At $t=3.0\,{\rm Gyr}$, magnetic field strength peaks around 
$x \sim 7$. 
This strength is about 80 times higher than the initial value. 
In the right panel, the minimum plasma $\beta$ appears at $x\sim 6$. 
In this region, the plasma $\beta$ is smaller than 10. 

\begin{figure}[ht]
\epsscale{.75}
\plotone{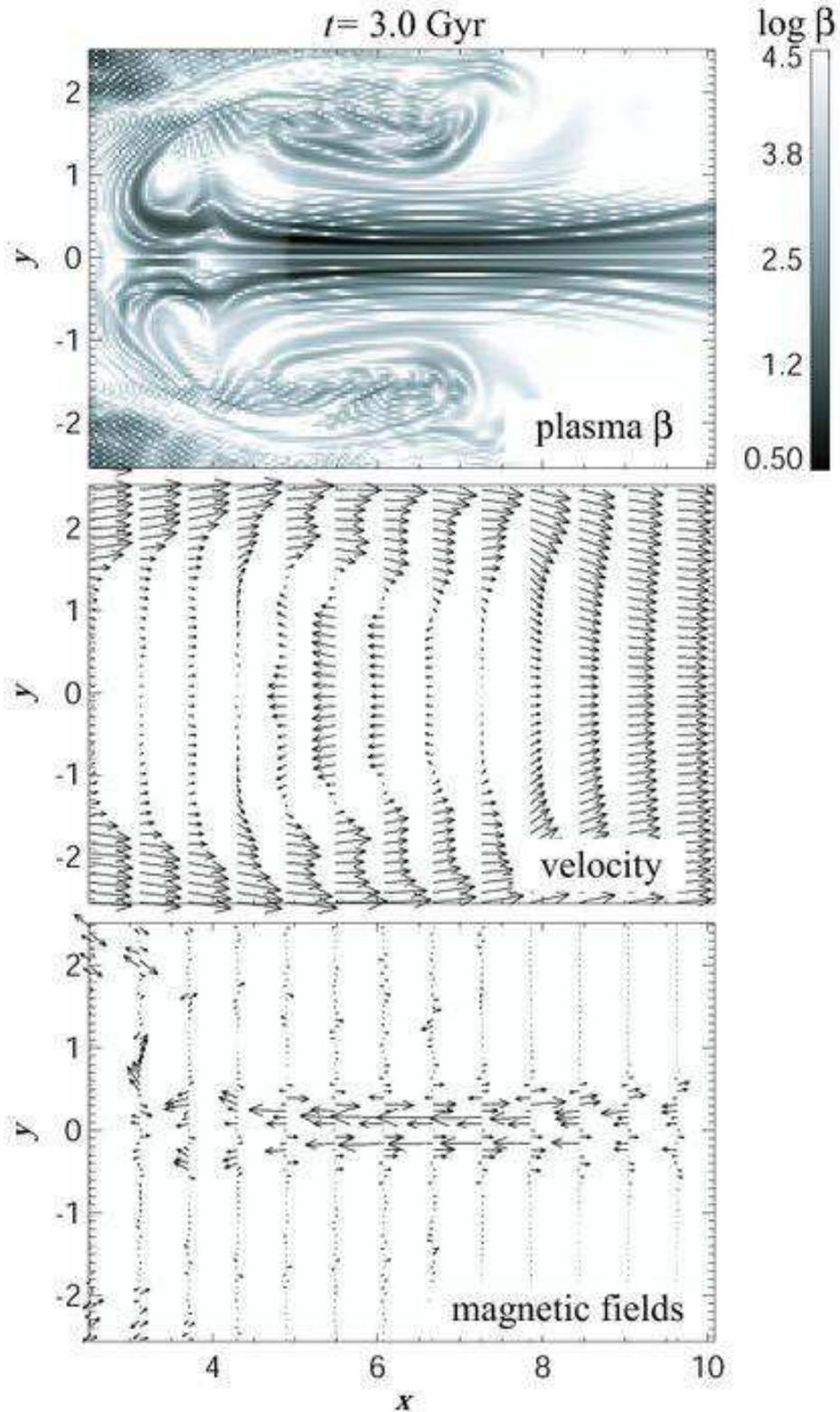}
\caption{Distributions of plasma $\beta$, velocity vectors, and magnetic
 field vectors from top to bottom in $z=0$ plane at $t = 3.0 \, {\rm
 Gyr}$ for model MT3. The region behind the subcluster 
$[x, \,y, \,z] = [2.5:10, \,-2.5:2.5, \,0]$ is shown. The front of the
 subcluster around $x\sim 0$ is not shown here. Arrows in middle and
 bottom panels show velocity vectors and magnetic field vectors.
\label{fig9}}
\end{figure}

Figure \ref{fig9} shows the distributions of plasma $\beta$, velocity
vectors, and magnetic field vectors in $z=0$ plane at 
$t = 3.0 \, {\rm Gyr}$ for model MT3. 
The region behind the subcluster, $2.5 \leq x \leq 10$, $-2.5 \leq y
\leq 2.5$, $z=0$ is shown.
The front of the subcluster is not shown here 
(it is located at $x\sim 0$). The top panel shows that $\beta \la 10$
along the $x$-axis ($y=z=0$).
We found that the ambient gas passing through the subcluster generates
vortices behind the subcluster. They create back flows along the
$x$-axis (see middle panel). The back flow is similar to that reported
by \citet{hei03}. Magnetic fields dragged by the gas flow are thus
stretched along the $x$-axis behind the subcluster. The $x$-component of
magnetic fields is particularly strengthened in the vicinity of the
$x$-axis (see bottom panels).

\section{Discussion and Summary}\label{sum}
We carried out 3D MHD simulations of a subcluster moving through a
magnetically turbulent ICM in order to study the effects of magnetic
fields on thermal conduction. We assumed that a dense, cold subcluster
with a sharp discontinuity in density and temperature is moving with
sound speed. 

In this paper, we studied whether a cold front can be sustained in a
magnetically turbulent ICM. In \S \ref{res1} and \S \ref{res2}, we
demonstrated that a cold front is maintained for over $1\, {\rm Gyr}$
because magnetic fields stretched along the front suppress the thermal
conduction across the front even if magnetic fields are initially
turbulent. On the other hand, when magnetic fields do not exist, steep
temperature gradients cannot be maintained because the cold subcluster
is heated up due to the isotropic thermal conduction. Therefore, we
conclude that cold fronts in merging clusters can exist because magnetic
fields coupled with the motion of the subcluster suppress the thermal
conduction. 

In \S \ref{res3}, we showed that magnetic fields are amplified
significantly behind the subcluster because the ambient flow converges
to the tail of the subcluster. The flow accumulates magnetic fields to
this tail region. Furthermore, the vortex motions behind the subcluster
accumulates and stretches the magnetic fields. The enhanced fields are
maintained for a long time behind the subcluster. Thus, the motion of a
subcluster forms a long tail of ordered magnetic fields. 
It may be worth noting that the magnetic filaments created behind the
subcluster are similar to the magnetic filaments in the solar
atmospheres, in which the magnetic fields are accumulated in boundaries
of the convective cell. Plasma $\beta$ decreases to $\beta \la 10$. 
If such a long tail of magnetic fields interacts with other moving
subclumps, magnetic fields will be further amplified. Consequently,
small-scale weak magnetic fields in the ICM can be amplified and create
large-scale magnetic fields whose energy is comparable to the thermal
energy. This mechanism may also apply to the amplification of
small-scale primordial magnetic fields once dark matter clumps are
formed.

\acknowledgments
We thank T. Yokoyama for developments of the coordinated astronomical
numerical software (CANS) which include 2D and 3D MHD codes including
thermal conduction. The development of CANS was supported by ACT-JST of
Japan Science and Technology Corporation.This work is supported by JSPS
Research Fellowships for Young Scientists. Numerical computations were
carried out on VPP5000 at the Center for Computational Astrophysics,
CfCA, of the National Astronomical Observatory of Japan and joint
research program of IMIT, Chiba University.

\clearpage
\begin{table}[ht]
\begin{center}
\caption{Simulation models and parameters. \label{tbl-1}}
\begin{tabular}{ccccccc}
\tableline\tableline
Model &$\kappa$\tablenotemark{a} & magnetic field & $B_{0} [\mu {\rm G}]$\tablenotemark{b} 
& $\beta_{0}$\tablenotemark{c} & box size [${\rm Mpc}^{3}$] & number of grids \\
\tableline
MT1 & $\kappa_{\parallel}$ & turbulent & $0.03 $ & $7.5 \times 10^{4}  $ &$2.5^{3} $ & $256^{3}$ \\
MT2 & $\kappa_{\parallel}$ & turbulent & $0.09  $ & $8.3 \times 10^{3}$ &$2.5^{3} $ & $256^{3}$ \\
MU1 & $\kappa_{\parallel}$ & uniform   & $0.07  $ & $1.0 \times 10^{4} $ &$2.5^{3} $ & $256^{3}$ \\
MU2 & $\kappa_{\parallel}$ & uniform   & $0.27  $ & $1.0 \times 10^{3} $ &$2.5^{3} $ & $256^{3}$ \\
MT3 & 0                    & turbulent & $0.03 $ & $7.5 \times 10^{4}  $ &$3.75\times 2.5\times 2.5$ & $384\times 256\times 256$ \\
H  & $\kappa$              & ---       & $ 0   $ & $\infty$            &$2.5^{3} $ & $256^{3}$ \\
\tableline
\end{tabular}
\tablenotetext{a}{$\kappa$ is the thermal conductivity, and the
 subscript $\parallel$ denotes the component parallel to magnetic field
 lines.}
\tablenotetext{b,c}{$B_{0}$ and $\beta_{0}$ are initial mean magnetic
 field strength and initial mean plasma $\beta$.}
\end{center}
\end{table}

\end{document}